\documentclass[sigconf]{acmart}
\AtBeginDocument{%
  \providecommand\BibTeX{{%
    \normalfont B\kern-0.5em{\scshape i\kern-0.25em b}\kern-0.8em\TeX}}}

\setcopyright{acmcopyright}
\copyrightyear{2025}
\acmYear{2025}
\acmDOI{XXXXXXX.XXXXXXX}

\acmConference[Conference acronym 'XX]{Make sure to enter the correct
  conference title from your rights confirmation email}{June 03--05,
  2018}{Woodstock, NY}
\acmPrice{15.00}
\acmISBN{978-1-4503-XXXX-X/18/06}

\usepackage{hyperref}
\usepackage{cleveref}
\usepackage{subcaption}
\usepackage{enumitem}
\usepackage{algorithm}
\usepackage{algpseudocode}
\usepackage{amsmath}
\usepackage{bbm}
\begin{document}

\title{Hierarchical Group-wise Ranking Framework for Recommendation Models}

\author{YaChen Yan}
\email{yachen.yan@creditkarma.com}
\orcid{0000-0002-1213-4343}
\affiliation{%
    \institution{Credit Karma}
    \streetaddress{760 Market Street}
    \city{San Francisco}
    \state{California}
    \country{USA}
    \postcode{94012}
}

\author{Liubo Li}
\email{liubo.li@creditkarma.com}
\orcid{1234-5678-9012}
\affiliation{%
    \institution{Credit Karma}
    \streetaddress{760 Market Street}
    \city{San Francisco}
    \state{California}
    \country{USA}
    \postcode{94012}
}

\author{Ravi Choudhary}
\email{ravi.choudhary@creditkarma.com}
\orcid{1234-5678-9012}
\affiliation{%
    \institution{Credit Karma}
    \streetaddress{760 Market Street}
    \city{San Francisco}
    \state{California}
    \country{USA}
    \postcode{94012}
}

\renewcommand{\shortauthors}{YaChen, LiuBo and Ravi}
\renewcommand{\subtitle}{Hierarchical Group-wise Ranking Framework for Recommendation Models}


\begin{abstract}
In modern recommender systems, CTR/CVR models are increasingly trained with ranking objectives to improve item ranking quality. While this shift aligns training more closely with serving goals, most existing methods rely on in-batch negative sampling, which predominantly surfaces easy negatives. This limits the model's ability to capture fine-grained user preferences and weakens overall ranking performance. To address this, we propose a Hierarchical Group-wise Ranking Framework with two key components. First, we apply residual vector quantization to user embeddings to generate hierarchical user codes that partition users into hierarchical, trie-structured clusters. Second, we apply listwise ranking losses to user-item pairs at each level of the hierarchy, where shallow levels group loosely similar users and deeper levels group highly similar users, reinforcing learning-to-rank signals through progressively harder negatives. Since users with similar preferences and content exposure tend to yield more informative negatives, applying ranking losses within these hierarchical user groups serves as an effective approximation of hard negative mining. Our approach improves ranking performance without requiring complex real-time context collection or retrieval infrastructure. Extensive experiments demonstrate that the proposed framework consistently enhances both model calibration and ranking accuracy, offering a scalable and practical solution for industrial recommender systems.
\end{abstract}



\maketitle

\section{Introduction}
Click-through rate (CTR) and conversion rate (CVR) prediction models play a pivotal role in large-scale recommender systems and online advertising. While most modern systems rely on binary classification objectives such as log loss to estimate the likelihood of user actions, enhancing the ranking quality of model predictions has become a critical direction for improving user experience and achieving business goals. In this context, learning-to-rank (LTR) objectives, including pairwise and listwise losses, are widely adopted to better capture users’ relative preferences among items.

However, a persistent challenge lies in constructing meaningful item comparisons during training. In particular, existing ranking losses often rely on in-batch negative sampling or uniformly sampled negative pairs, which tend to overemphasize easy negatives while underutilizing more informative, harder negatives. Recent research has shown that sampling negatives based on similarity or gradient-based importance can significantly improve model performance, but often at the cost of increased computational overhead, particularly in real-time environments. Existing context-aware approaches such as JRC \cite{sheng2023joint} and SBCR \cite{zhang2024self} improve ranking performance by leveraging online ranked list logging. However, these methods require real-time infrastructure and tightly integrated systems, which increases deployment complexity and limits scalability in production environments. Furthermore, CVR models often suffer from sparse in-session user feedback, limiting the effectiveness of context-aware negative sampling based on in-session interactions.

To address these challenges, we propose a novel \textbf{Hierarchical Group-wise Ranking Framework} that improves ranking performance without relying on real-time context or nearest-neighbor retrieval. Our approach uses residual vector quantization (RVQ) to learn hierarchical user codes and group user-item pairs into multi-level clusters. Within each group, we apply listwise ranking losses over progressively harder negatives, based on the intuition that users with similar profiles or behaviors yield more informative comparisons. This hierarchical, multi-resolution cross-user sampling provides an efficient and scalable alternative for industrial recommendation systems. The main contributions of this paper can be summarized as follows:

\begin{itemize}
    \item We propose a residual vector quantization module to encode user embeddings into hierarchical discrete codes, which serve as the foundation for multi-level user grouping. This structure enables dynamic and granular control over the difficulty level of sampled negatives during training.
    \item We introduce a hierarchical group-wise listwise ranking loss that applies ranking loss within user groups defined at each hierarchical level. By varying the granularity level of grouping, our method progressively surfaces harder negatives, offering an efficient alternative to gradient-based sampling strategies.
    \item We integrate this hierarchical ranking objective with standard calibration losses in a multi-task learning framework and demonstrate that our method improves both convergence efficiency and ranking performance across multiple domains, without requiring real-time context collection or retrieval infrastructure.
    \item We conduct extensive experiments on real-world datasets to analyze the performance and training dynamics of the proposed hierarchical group-wise ranking framework and demonstrate its effectiveness.
\end{itemize}

\section{Related Work}

\subsection{Integrating Ranking Objective into CTR/CVR Prediction}
Researchers have proposed combining binary logistic loss (logloss) with auxiliary ranking losses to jointly optimize calibration accuracy and ranking quality in recommendation systems. Twitter pioneered this approach by integrating logloss with pairwise ranking loss for CTR prediction in its timeline \cite{li2015click}, demonstrating substantial gains across key metrics. This multi-objective framework has since been widely adopted by major platforms, including Google \cite{yan2022scale, bai2023regression} and Alibaba \cite{sheng2023joint}. Calibrated Softmax \cite{yan2022scale} addressed the challenge of calibration in deep ranking models, showing that conventional ranking losses such as softmax can lead to score instability during training. They proposed a multi-objective framework that maintains strong ranking performance while producing calibrated outputs, laying the foundation for later calibrated ranking approaches. Building on this, Regression-Compatible Ranking (RCR) \cite{bai2023regression} aligned logloss with a modified listwise loss to enable compatible optimization. Joint Ranking and Calibration (JRC) \cite{sheng2023joint} introduced a dual-logit architecture to separately model click and non-click states. The Self-Boosted Framework for Calibrated Ranking (SBCR) \cite{zhang2024self} leveraged ranking scores from deployed models as contextual features, allowing for extensive data shuffling while preserving ranking quality. This decoupled the calibration and ranking objectives through distinct optimization modules. A gradient-based explanation by \cite{lin2024understanding} showed that ranking loss mitigates gradient vanishing in negative samples when positive feedback is sparse, thereby improving both classification accuracy and ranking performance. This theoretical insight helps unify prior empirical observations.

\subsection{Vector Quantization}
Vector Quantization (VQ) has emerged as a promising technique across various machine learning domains, particularly in recommendation systems \cite{liu2024vector}. VQ-VAE \cite{van2017neural} combines vector quantization with variational autoencoders to learn discrete latent representations, demonstrating effectiveness in image, audio, and video generation tasks. The technique has been further extended in audio processing by SoundStream \cite{zeghidour2021soundstream}, an end-to-end neural audio codec that leverages residual vector quantization to achieve high-quality compression at low bitrates, showcasing VQ's cross-modal versatility. In the image domain, RQ-VAE \cite{lee2022autoregressive} employs residual quantization for autoregressive image generation, showing that hierarchical quantized representations enable efficient high-fidelity synthesis without adversarial training. In recommendation systems, TIGER \cite{rajput2023recommender} introduced a generative retrieval framework that uses semantic IDs derived from content embeddings via quantization schemes such as RQ-VAE. This approach significantly outperforms traditional recommendation models by creating hierarchical item representations that improve cold-start performance and result diversity. The authors further demonstrated that hierarchical semantic IDs can replace traditional item IDs in large-scale recommender systems, enhancing generalization. Streaming VQ \cite{bin2025real} has also been applied to real-time indexing, dynamically assigning items to clusters while maintaining balanced index distributions. Unlike conventional methods that require periodic index rebuilding, streaming VQ supports continuous item indexing without service disruption, allowing real-time adaptation to evolving user preferences and content dynamics.

\section{Preliminaries}
The recommendation model with binary relevance is commonly formulated as a binary classification problem optimized with binary logistic loss. The binary logistic loss function is defined as:
\begin{equation}
\mathcal{L}_{\text{logloss}} = -\frac{1}{N}\sum_{i=1}^{N}[y_i\log(\hat{y}_i) + (1-y_i)\log(1-\hat{y}_i)]
\end{equation}
where $\hat{y}_i$ represents the predicted probability and $y_i \in \{0, 1\}$ is the binary feedback label. To enhance ranking performance, recent approaches incorporate Learning-to-Rank (LTR) losses as auxiliary objectives. The combined objective function $\mathcal{L} = \mathcal{L}_{\text{logloss}} + \lambda\mathcal{L}_{rank}$ integrates binary logistic loss with either pairwise ranking loss (comparing item pairs) or listwise ranking loss (optimizing the entire item list ordering), where $\lambda$ controls the contribution of the ranking component. In this setting, the ranking loss typically operates on the model’s predicted scores over user-item pairs. Negative samples for the ranking loss are commonly drawn uniformly from other user-item pairs within the same training batch (i.e., in-batch negative sampling) to construct contrastive comparisons during training.

Without loss of generality, the notation used throughout this paper for recommendation models with binary relevance is as follows: the training dataset consists of instances $(x_{u}, x_{i}, y)$, where $x_{u}$ and $x_{i}$ denote user and item features respectively, $y \in \{0, 1\}$ indicates the binary user-item feedback label. The raw user features $x_u$ and item features $x_i$ are processed by their respective networks to obtain user embedding $\mathbf{e}_u$ and item embedding $\mathbf{e}_i$. These embeddings are then fed into a main network that outputs logit $s$, which is transformed into the predicted probability $\hat{y} = \sigma(s)$ using the sigmoid function $\sigma$. The primary optimization objective is the binary logistic loss comparing $\hat{y}$ against ground truth labels $y$, while a ranking loss encourages correct ordering of items. This general notation serves as the foundation for our proposed model as well, which extends these concepts with additional specialized notation to capture our proposed framework.

\section{Negative Sampling Challenges and Theoretical Gradient Analysis}
Despite the effectiveness of combining binary logistic loss with ranking loss for CTR/CVR prediction, conventional approaches still face significant optimization challenges. In this section, we analyze two critical limitations of existing ranking losses used in CTR/CVR models: the inefficiency of traditional negative sampling strategies and their suboptimal convergence properties.

Our analysis reveals that conventional sampling techniques often focus on easy negatives that contribute little to model refinement. Through analyzing gradient dynamics, we show why sampling harder negative examples in proportion to their gradient norms can improve convergence. This analysis motivates our hierarchical group-wise sampling approach, which systematically surfaces challenging negative examples at varying difficulty levels from similar users, without requiring computationally expensive nearest-neighbor searches.

\subsection{Limitations of Uniform Sampling}
Traditional negative sampling strategies, such as uniform in-batch sampling, often prioritize easy negatives that are trivial for the model to distinguish. This limits the model’s ability to learn fine-grained user preference signals and ultimately undermines ranking performance. To address this, recent work has explored retrieving harder negatives via approximate nearest neighbor (ANN) search \cite{xiong2020approximate}, which improves the model’s capacity to capture subtle distinctions. However, ANN-based methods introduce significant computational overhead and are generally better suited to retrieval models than to ranking models, which require capturing complex user-item interactions beyond inner products and supporting online learning. The key challenge is to efficiently surface informative hard negatives during training, without relying on real-time context collection or exhaustive nearest-neighbor search.

\subsection{Gradient-Based Sampling Theory}
Consider a training batch with positive and negative samples. An importance-weighted stochastic gradient descent (SGD) update for the ranking loss can be expressed as:
\begin{equation}
    \theta_{t+1} = \theta_{t} - \eta\frac{1}{Np^{-}}\nabla_{\theta_t}l(s^+, s^-)
\end{equation}
where $\eta$ is the learning rate. $\theta_t$ represents the parameters at iteration $t$, $\theta_{t+1}$ the updated parameters. $p^{-}$ denotes the probability of selecting a particular negative instance $(x_u, x_i,y^-)$. The scaling factor $\frac{1}{N p^{-}}$ ensures an unbiased gradient estimate. Following derivations in variance reduction \cite{katharopoulos2018not, johnson2018training}, let $g = \frac{1}{Np^{-}} \nabla_{\theta_t}l(s^+, s^-)$ be the weighted gradient, we can write the convergence rate as:
\begin{equation}
\begin{aligned}
    E\Delta_t & = \|\theta_t - \theta^*\|^2 - E_{P^{-}}(\|\theta_{t+1} - \theta^*\|^2) \\
    &= 2\eta E_{P^{-}}(g)^T(\theta_t - \theta^*) - \eta^2 E_{P^{-}}(g)^T E_{P^{-}}(g) \\
    & \quad - \eta^2 \text{Tr}(V_{P^{-}}(g))
\end{aligned}    
\end{equation}
where $P^{-}$ is the negative sampling distribution for a given positive example $(x_u, x_i, y^+)$. This formulation shows that convergence can be improved by selecting negative examples from a distribution that reduces $\text{Tr}(V_{P^{-}}(g))$, which quantifies the total gradient variance introduced by the negative sampling. The optimal sampling strategy is
\begin{equation}
    p^{*-} = \arg\min_{p^{-}} \text{Tr}(V_{P^{-}}(g)) \propto \|\nabla_{\theta_t}l(s^+, s^-)\|^2
\end{equation}
The above analysis shows that the optimal negative sampling distribution is proportional to the squared gradient norm of each instance, favoring samples that contribute larger updates to the model. Intuitively, a negative instance with a larger gradient norm is more likely to reduce the training loss, while those with vanishing gradients tend to be less informative. This aligns with established variance reduction principles in stochastic optimization and motivates a gradient-informed sampling strategy.

This approach is particularly valuable in recommendation systems, where the vast majority of negative examples contribute minimal learning signals. The resulting long-tail distribution of negatives can dramatically slow model training when not properly addressed through strategic sampling techniques. Prioritizing hard negatives with large gradient contributions reduces the variance of gradient estimates and accelerate convergence.

\section{Proposed Framework}
To operationalize the above theoretical insight, we propose a hierarchical group-wise negative sampling strategy that approximates the optimal distribution without incurring costly computations. Our approach clusters users based on profile or behavioral similarity, and groups the associated user-item samples according to these user clusters across multiple hierarchical levels. This structure enables the model to sample negatives from users of varying similarity: from coarse groups capturing general behavior patterns to fine-grained subgroups reflecting closely shared interests. Intuitively, negatives samples drawn from similar users are more informative, as similar users tend to be exposed to overlapping content and exhibit comparable preferences. By leveraging negative samples from users with aligned exposure and interests, our method introduces progressively harder negatives that enhance the learning signal throughout the hierarchy. This promotes more effective optimization of ranking loss while maintaining computational efficiency, ultimately improving learning dynamics in CTR/CVR prediction tasks.

\begin{figure}[htpb]
    \centering
    \includegraphics[width=0.50\textwidth, height=0.35\textwidth]{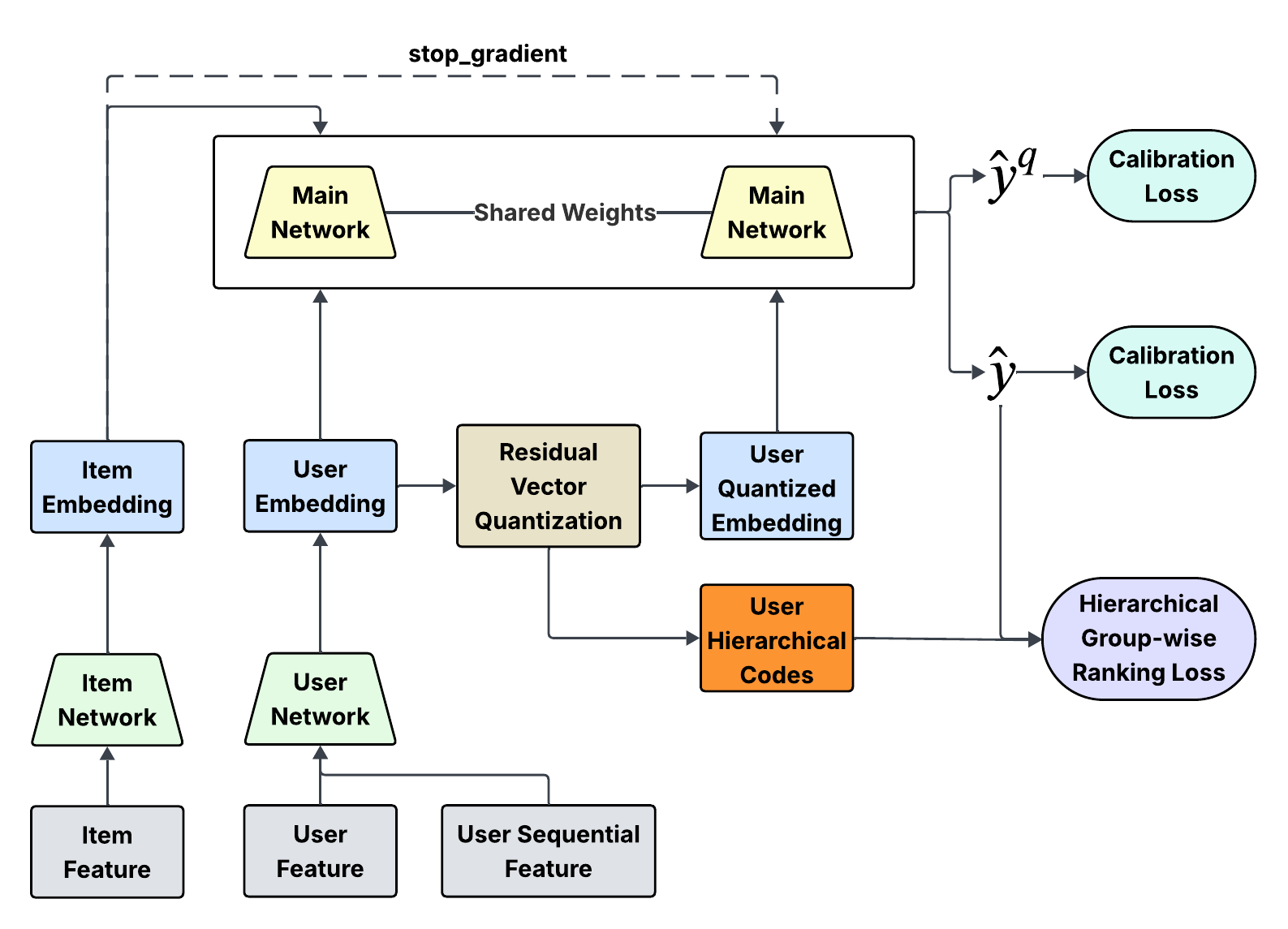}
    \caption{The Architecture of the Proposed Framework}
    \label{fig_architecture}
    \medskip
    \small
\end{figure}

We designed the proposed framework, as illustrated in \autoref{fig_architecture} with three main components:

\begin{itemize}
    \item \textbf{Hierarchical User Code Generation}: We quantize each user embedding into a structured sequence of discrete codes using multi-stage residual vector quantization, where each stage uses a codebook to quantize the remaining error from the previous level. This produces a hierarchical code sequence that forms a trie-like structure, where higher levels represent broad semantic groupings and deeper levels capture fine-grained user distinctions.
    \item \textbf{Hierarchical Group-wise Ranking Objective}: Based on the generated user codes, we organize users into nested groups where users sharing the same prefix codes at each level are grouped together, forming a trie-like structure of increasing similarity. We apply listwise ranking loss within each group, computing the loss over groups containing user-item pairs with shared code prefixes. By varying the group depth, we control negative difficulty: shallow levels provide easier negatives from loosely similar users, while deeper levels yield harder negatives from highly similar users. To balance contributions across hierarchy levels, we employ an uncertainty-based weighting scheme, enabling the model to adaptively focus on the most informative hierarchy depths during training.
    \item \textbf{Multi-Objective Training Strategy}: Our training objective combines three components: a primary calibration loss on predictions from the original user embedding, an auxiliary calibration loss on predictions from the quantized embedding (using a straight-through estimator to enable gradient flow), and the proposed hierarchical ranking loss. Unlike traditional vector quantization approaches, we omit commitment loss to preserve adaptability in dynamic recommendation settings, instead relying on the auxiliary calibration loss to encourage alignment between original and quantized embeddings while maintaining flexibility for evolving user preferences and behaviors.
\end{itemize}

In our model specifically, the user embedding $\mathbf{e}_u$ undergoes residual vector quantization to produce the quantized user embedding $\mathbf{e}_{u}^{q}$ and user hierarchical codes $\mathbf{c}_u$. When the shared main network uses the quantized user embedding $\mathbf{e}_{u}^{q}$ and the item embedding $\mathbf{e}_i$ as inputs, it produces logit $s^{q}$ and the corresponding predicted probability $\hat{y}^{q} = \sigma(s^{q})$. To avoid conflicting gradients between the dual prediction paths, we stop the gradient flow from the auxiliary loss into the item embedding by applying $\text{stop\_gradient}(\mathbf{e}_i)$ when computing $\hat{y}^{q}$. This ensures that only the user network receives updates from the auxiliary calibration loss, preserving training stability. Both the original and quantized predictions ($\hat{y}$ and $\hat{y}^{q}$) are optimized using calibration loss against ground truth labels. Additionally, the user hierarchical codes $\mathbf{c}_u$ are used to compute a hierarchical group-wise ranking loss to further enhance ranking performance. These notations ($\mathbf{e}_{u}^{q}$, $\mathbf{c}_u$, $s^{q}$, $\hat{y}^{q}$) extend the general framework presented earlier and are specific to our quantization-based approach, introducing additional variables necessary to describe our model's unique architecture and optimization strategy.

\subsection{Hierarchical User Codes Generation}

\begin{figure}[htpb]
    \centering
    \includegraphics[width=0.50\textwidth, height=0.25\textwidth]{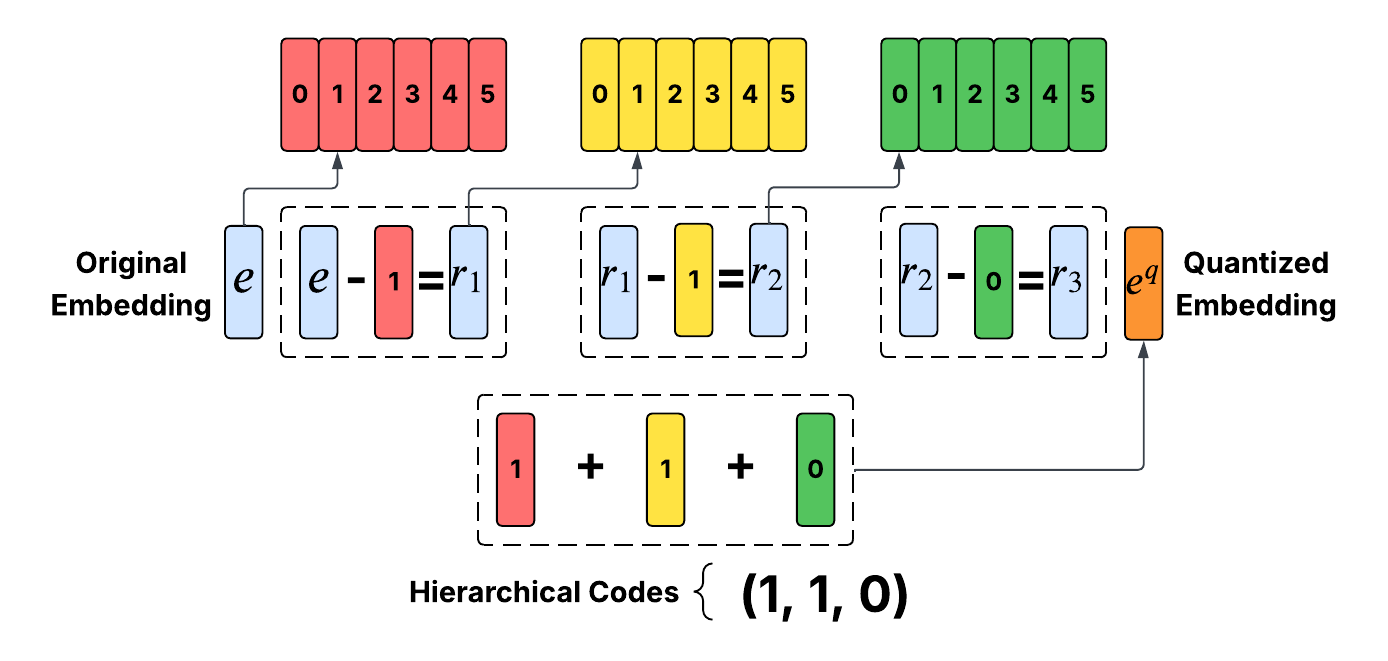}
    \caption{Residual Vector Quantization}
    \label{fig_rvq}
    \medskip
    \small
\end{figure}

To capture structured user similarity and enable efficient group-wise sampling, we discretize user embeddings using a residual vector quantization (RVQ) framework. This process encodes each user into a sequence of discrete codes, referred to as \textbf{hierarchical user codes}, which form the foundation of our multi-resolution user grouping strategy. \autoref{fig_rvq} illustrates the cascaded quantization process and the resulting hierarchical user code structure.

Given a user embedding $\mathbf{e}_u \in \mathbb{R}^d$ produced by the user network (which may include contextual or sequential features), we apply an $L$-stage residual quantization procedure to obtain a code sequence $\mathbf{c}_u = [\mathbf{c}_{u,1}, \ldots, \mathbf{c}_{u,L}]$. At each stage $l$, a codebook $\mathcal{C}^{(l)} = \{\mathcal{C}^{(l)}_1, \ldots, \mathcal{C}^{(l)}_K\}$ is used to quantize the residual vector passed down from the previous level. These codebooks are arranged in a cascaded structure, where each level incrementally refines the remaining quantization error:
\begin{equation}
\begin{aligned}
    \mathbf{r}_u^{(1)} &= \mathbf{e}_u \\
    \mathbf{c}_{u,l} &= \arg\min_{k} \|\mathbf{r}_u^{(l)} - \mathcal{C}^{(l)}_k\|_2^2 \\
    \mathbf{r}_u^{(l+1)} &= \mathbf{r}_u^{(l)} - \mathcal{C}^{(l)}_{\mathbf{c}_{u,l}}
\end{aligned}
\end{equation}
The quantized embedding is reconstructed by summing codebook vectors from all stages:
\begin{equation}
    \hat{\mathbf{e}}_u = \sum_{l=1}^{L} \mathcal{C}^{(l)}_{\mathbf{c}_{u,l}}
\end{equation}

To ensure stable learning and effective usage of codebook entries, each codebook is updated using Exponential Moving Average (EMA) strategy. Following each assignment, usage statistics and accumulated residuals are used to softly update the code vectors:
\begin{equation}
    \mathcal{C}^{(l)}_k \leftarrow m \cdot \mathcal{C}^{(l)}_k + (1 - m) \cdot \mu_k,\quad
    N_k \leftarrow m \cdot N_k + (1 - m) \cdot n_k
\end{equation}
where $\mu_k$ and $n_k$ represent the average and count of residuals assigned to code $k$, and $m$ is the EMA decay rate. We also apply Laplace smoothing to the EMA count to avoid instability from rare updates.

To prevent underutilized or inactive codes, we include an expiration mechanism that replaces codes with usage counts below a predefined threshold using randomly sampled embeddings from the current batch, following SoundStream~\cite{zeghidour2021soundstream}. This promotes codebook diversity and avoids representation collapse during training.

The resulting hierarchical code $\mathbf{c}_u = [\mathbf{c}_{u,1}, \ldots, \mathbf{c}_{u,L}]$ defines a multi-level user grouping scheme, where each user is assigned a path through a cascade of discrete decisions. Intuitively, we can think of hierarchical codes as forming a trie over users, with a coarse-to-fine semantic structure. Users sharing longer prefix codes occupy the same subtree and are thus considered semantically closer. As we advance to deeper levels in the hierarchy, the granularity of user grouping increases, enabling fine-grained discrimination among users with otherwise similar high-level behavior. This structure naturally enables efficient, multi-resolution user grouping: sampling at higher levels yields broad diversity, while sampling at lower levels targets hard negatives from similar users.

\subsection{Hierarchical Group-wise Ranking Objective}
The hierarchical user code $\mathbf{c}_u = [\mathbf{c}_{u,1}, \ldots, \mathbf{c}_{u,L}]$ assigns each user to a sequence of discrete code indices, embedding them into a multi-resolution user group. At each level $l$, users sharing the same prefix $(\mathbf{c}_{u,1}, \ldots, \mathbf{c}_{u,l})$ are grouped together, forming nested user clusters of increasing similarity. This trie-like structure enables us to define semantically coherent user-item groups at varying levels of granularity, where deeper levels reflect finer user proximity.

\begin{figure*}
\centering
\includegraphics[width=0.80\textwidth, height=0.25\textwidth]{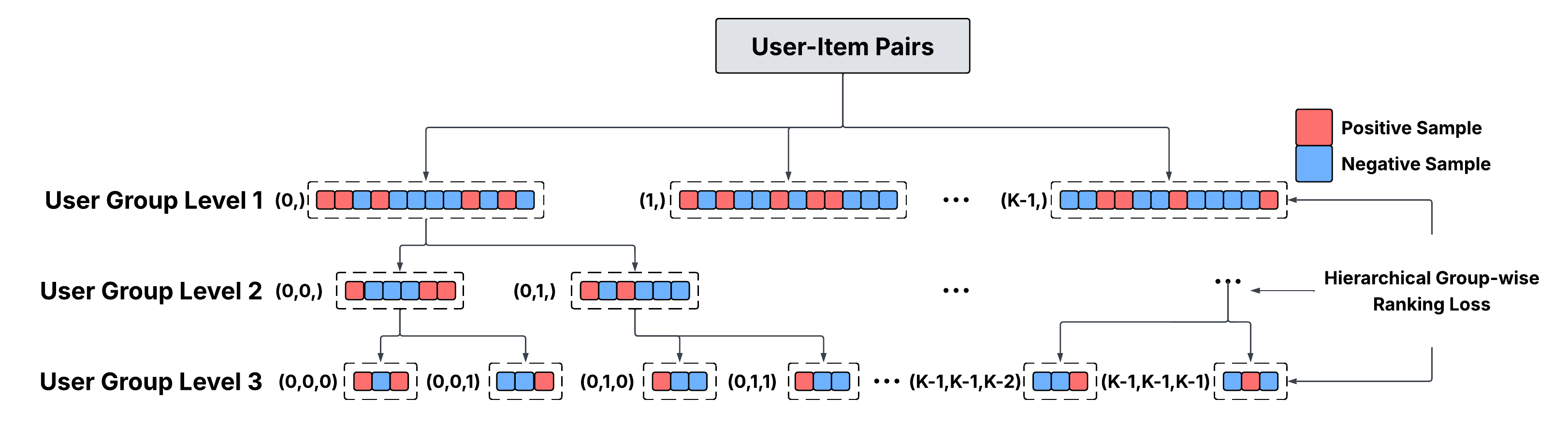}
\caption{Hierarchical Group-wise Ranking Framework. A trie-structured approach organizes user-item pairs into multi-level user groups based on shared code prefixes. Positive (red) and negative (blue) examples are drawn within each group to supervise ranking objectives. Varying the group depth enables learning to rank over increasingly fine-grained user similarity.}
\label{fig_hierarchy}
\medskip
\small
\end{figure*}

As illustrated in \autoref{fig_hierarchy}, we recursively partition user-item pairs into finer groups according to these hierarchical prefixes. Within each group, the users' positive items are treated as positive examples, while users' negative items serve as negatives. By varying the group depth $l$, we effectively control the difficulty of sampled negatives: shallower levels provide easier negatives from loosely similar users, while deeper levels yield harder negatives from highly similar users who share more behavioral or contextual overlap and content exposure.

To train the model on these grouped samples, we adopt the Regression Compatible Listwise Cross Entropy loss (ListCE)\cite{bai2023regression}, which replaces the softmax transformation with the sigmoid function based normalization. This improves compatibility between ranking and calibration losses under binary relevance.

Let $G^{(l)}_1, G^{(l)}_2, \ldots, G^{(l)}_{M_l}$ denote the groups formed at level $l$, where each group $G^{(l)}_m$ contains user-item pairs sharing the same code prefix. The listwise loss at level $l$ is defined as:

\begin{equation}
\mathcal{L}^{(l)}_{\text{listce}}(s, y) = \frac{1}{M_l} \sum_{m=1}^{M_l}
\sum_{i \in G^{(l)}_m}
- \tilde{y}_i^{(l,m)} \log \left( \frac{\sigma(s_i)}{\sum_{j \in G^{(l)}_m} \sigma(s_j)} \right)
\end{equation}

where $s_i$ is the predicted logit for user-item pair $i$, and $\sigma(s_i)$ represents its sigmoid-transformed score. The corresponding normalized label $\tilde{y}_i^{(l,m)}$ is computed as: $\tilde{y}_i^{(l,m)} = \frac{y_i}{\sum_{j \in G^{(l)}_m} y_j + \epsilon}$,
ensuring that labels are normalized within each group $G^{(l)}_m$.

This loss function achieves global minima when the sigmoid-normalized scores match the soft label distribution, thereby aligning the ranking objective with the binary classification goal. Compared to softmax-based listwise losses, ListCE avoids the scale-misalignment issue and facilitates simultaneous optimization with the log loss calibration objective.

To balance the contribution from different hierarchy levels, we introduce an uncertainty-based weighting scheme \cite{kendall2018multi}. Each level $l$ is associated with a learnable uncertainty parameter $\sigma_l$. The total hierarchical ranking loss is defined as:

\begin{equation}
\mathcal{L}_{\text{hierarchical}} = \sum_{l=1}^{L} \left( \frac{1}{2\sigma_l^2} \mathcal{L}^{(l)}_{\text{listwise}}(s, y) + \log \sigma_l \right)
\end{equation}

This formulation enables the model to adaptively weight ranking loss from each level based on its estimated uncertainty. As a result, the model learns to focus on the most informative hierarchy depths during training, while preserving stable and balanced optimization across levels.

\subsection{Multi-Objective Training Strategy}
\subsubsection{Objective Function Formulation}
Our overall training objective integrates three components: a primary calibration loss on the predicted click-through probability from the original user embedding, a secondary calibration loss from the quantized user embedding, and a hierarchical group-wise listwise ranking loss. Formally, the total loss is defined as:

\begin{equation}
\begin{aligned}
    \mathcal{L}_{\text{loss}}  & = \mathcal{L}_{\text{logloss}}(\hat{y}, y) \\
    & + \lambda\mathcal{L}_{\text{logloss}}(\hat{y}^{q}, y) \\
    & + \mathcal{L}_{\text{hierarchical}}
\end{aligned}
\end{equation}

The first term, $\mathcal{L}_{\text{logloss}}(\hat{y}, y)$, serves as the primary loss for calibrating the model’s click-through probability prediction using the original user embedding $\mathbf{e}_u$. This component ensures that the model produces well-calibrated probability estimates suitable for real-world serving, ensuring compatibility with recommendation systems that consume predicted probabilities.

The second term, $\mathcal{L}_{\text{logloss}}(\hat{y}^{q}, y)$, introduces an auxiliary calibration loss applied to predictions derived from the quantized user embedding $\mathbf{e}_u^q$, obtained via residual vector quantization. Importantly, to enable backpropagation through the non-differentiable quantization operation, we apply a straight-through estimator (STE):

\begin{equation}
\mathbf{e}_{u}^{q} = \mathbf{e}_u + \text{stop\_gradient}(\hat{\mathbf{e}}_u - \mathbf{e}_u)
\end{equation}

Here, $\hat{\mathbf{e}}_u = \sum_{l=1}^{L} \mathcal{C}^{(l)}_{\mathbf{c}_{u,l}}$ represents the quantized embedding reconstructed from the multi-level codebooks. This formulation ensures that $\mathbf{e}_{u}^{q}$ behaves as the quantized embedding $\hat{\mathbf{e}}_u$ in the forward pass, while the backward pass treats it as if it were the original embedding $\mathbf{e}_u$. As a result, gradients bypass the discrete code selection path and directly update $\mathbf{e}_u$, allowing the user encoder to remain trainable while benefiting from quantization-based regularization.

Although $\mathbf{e}_u^q$ is not used during serving, this auxiliary loss regularizes the shared user network by encouraging it to produce embeddings that are not only predictive but also structurally compressible and cluster-aware. Specifically, it helps align the user embedding space with the quantized codebook space, promoting smoother transitions and more stable quantization behavior. This alignment facilitates better codebook utilization and supports dynamic clustering under shifting user distributions. As a result, it enhances the generalization and calibration quality of the primary prediction $\hat{y}$ and improves training stability for the hierarchical group-wise ranking objective by maintaining a semantically meaningful and adaptive latent structure.

The final component, $\mathcal{L}_{\text{hierarchical}}$, is the proposed hierarchical group-wise listwise ranking loss. It utilizes trie-structured user codes generated via residual vector quantization to group user-item pairs into semantically coherent user clusters at multiple levels of granularity. By applying the ranking loss within these groups, the model receives user cluster-aware ranking optimization using progressively harder negatives drawn from similar users. This structure improves ranking quality without incurring the cost of real-time context collection or explicit nearest-neighbor retrieval.

\subsubsection{Clustering Adaptability}
Traditional vector quantization frameworks such as VQ-VAE~\cite{van2017neural} typically employ a commitment loss (e.g., $\|\mathbf{e}_u - \hat{\mathbf{e}}_u\|^2$) to explicitly align the continuous embedding with its quantized counterpart. However, we omit this component in our framework due to the dynamic nature of real-time recommendation settings, where user embeddings must continually adapt to evolving preferences, behaviors, and contextual signals.

Enforcing a commitment loss would constrain user embeddings to remain near static quantized representations, limiting their ability to transition across clusters and adapt in response to new interactions. Instead, we allow the primary calibration loss $\mathcal{L}_{\text{logloss}}(\hat{y}, y)$ to guide representation learning, maintaining expressiveness and adaptability. To softly encourage alignment between the embedding and its quantized version, we incorporate an auxiliary calibration loss $\mathcal{L}_{\text{logloss}}(\hat{y}^{q}, y)$ applied to the quantized prediction. This auxiliary loss acts as a task-driven regularizer, promoting semantically meaningful and discretization-friendly embeddings without rigid constraints.

Our approach supports flexible and generalizable representation learning under streaming or non-stationary conditions, aligning with insights from recent work on real-time indexing~\cite{bin2025real}, which similarly avoids commitment loss to preserve adaptability.

\section{Experiments}
In this section, we provide a comprehensive evaluation of our proposed framework, covering dataset details, evaluation metrics, baseline comparisons, and result analysis. Experiments conducted on two large-scale public datasets from the short video recommendation and e-commerce domains demonstrate that our hierarchical group-wise ranking framework significantly improves ranking model performance.

During experiments, we focus on evaluating the effectiveness of our proposed models and answering the following questions.

\begin{itemize}
    \item \textbf{Q1}: How does our proposed framework perform on ranking tasks? Is it effective and efficient in extremely high-dimensional and sparse data settings?
    \item \textbf{Q2}: How well does our framework handle user cold-start scenarios with limited interaction history? Can it maintain robust performance when user signals are sparse?
    \item \textbf{Q3}: How do different losses and hyperparameter settings influence the performance of our framework?
\end{itemize}

\subsection{Experiment Setup}

\subsubsection{Datasets}
We evaluate our proposed model using two publicly available real-world datasets commonly utilized in research: KuaiRand and Taobao. The data is randomly divided into three subsets: 70\% for training, 10\% for validation, and 20\% for testing. We applied stratified sampling to ensure that each user has positive samples in every data subset.

\begin{itemize}
    \item \textbf{KuaiRand\footnote{https://kuairand.com/}} is a recommendation dataset collected from the video-sharing mobile app Kuaishou.
    \item \textbf{Taobao\footnote{https://tianchi.aliyun.com/dataset/649}} is a Taobao E-commerce dataset released Alibaba.
\end{itemize}

\subsubsection{Evaluation Metrics}
We consider LogLoss, AUC and GAUC to evaluate the performance of the models.

\textbf{LogLoss} LogLoss is both our loss function and evaluation metric. It measures the average distance between predicted probability and true label of all the examples.

\textbf{AUC} Area Under the ROC Curve (AUC) measures the probability that a randomly chosen positive example ranked higher by the model than a randomly chosen negative example. AUC only considers the relative order between positive and negative examples. A higher AUC indicates better ranking performance.

\textbf{GAUC} Group AUC (GAUC) computes the AUC separately for each user and then aggregates the results using a weighted average based on the number of impressions per user. It evaluates the model’s ability to rank items correctly within each user’s candidate set, reflecting intra-user ranking quality. GAUC is considered more consistent with real-world online performance and serves as a key metric in production systems.

\subsubsection{Reproducibility}
We implement all the models using Tensorflow~\cite{abadi2016tensorflow}. The mini-batch size is 4096, and the embedding dimension is $\max(\lfloor \log_2(\text{cardinality})\rfloor * 2, 16)$ for all the features. For optimization, we employ AdamW~\cite{loshchilov2017decoupled} with a learning rate set to 0.0001 for all the models. Grid-search for each competing model's hyper-parameters is conducted on the validation dataset. The number of DNN layers is from 2 to 4. The number of neurons ranges from 64 to 512. For the residual vector quantizer, we vary the codebook size from 4 to 32 and the number of quantization layers from 1 to 4. All the models are trained with early stopping and are evaluated every 1000 training steps.

\subsection{Model Performance Comparison (Q1)}

\begin{table}[htpb]
    \caption{Performance Comparison of Different Ranking Objectives on KuaiRand and Taobao Datasets.}
    \label{tbl_objective_performance}
    \centering
    \resizebox{1.0\linewidth}{!}{
        \begin{tabular}{l|ccc|ccc}
            \hline
            \textbf{Objective} & \multicolumn{3}{c|}{\textbf{KuaiRand}} & \multicolumn{3}{c}{\textbf{Taobao}} \\
            & LogLoss & AUC & GAUC & LogLoss & AUC & GAUC \\
            \hline
            LogLoss & 0.5735 & 0.7510 & 0.6911 & 0.2011 & 0.6420 & 0.5708 \\
            LogLoss + PairwiseLogistic & 0.5723 & 0.7524 & 0.6921 & 0.2002 & 0.6435 & 0.5728 \\
            LogLoss + SoftmaxCE & 0.5727 & 0.7520 & 0.6920 & 0.2005 & 0.6428 & 0.5720 \\
            LogLoss + ListCE & 0.5709 & 0.7537 & 0.6932 & 0.1995 & 0.6443 & 0.5734 \\
            JRC & 0.5713 & 0.7533 & 0.6930 & 0.1993 & 0.6540 & 0.5732 \\
            \hline
            GroupCE (proposed) & \textbf{0.5681} & \textbf{0.7556} & \textbf{0.6953} & \textbf{0.1982} & \textbf{0.6556} & \textbf{0.5745} \\
            \hline
        \end{tabular}
    }
\end{table}

The overall performance of different losses is listed in \autoref{tbl_objective_performance}. We have the following observations in terms of objective function effectiveness:
\begin{itemize}[leftmargin=10pt]
    \item \textbf{LogLoss} serves as the baseline objective and yields the lowest performance across both datasets, demonstrating the limitations of using a pure calibration loss without ranking optimization.
    \item \textbf{LogLoss + PairwiseLogistic} and \textbf{LogLoss + SoftmaxCE} show consistent improvements over the baseline, highlighting the benefit of incorporating pairwise or listwise ranking losses into model training.
    \item \textbf{LogLoss + ListCE} and \textbf{JRC} achieve further gains, demonstrating that listwise ranking objectives with calibration-compatible designs lead to stronger overall performance.
    \item \textbf{GroupCE} achieves the best performance across all metrics. These results validate the effectiveness of our hierarchical group-wise ranking strategy, which enables progressively harder negative sampling through structured user clustering.
\end{itemize}

\subsection{Cold Start Capability (Q2)}

To assess the model’s robustness in user cold-start scenarios, we evaluate its performance on user cohorts with limited interaction history. Since our model leverages hierarchical user codes for structured contrastive learning, we hypothesize that it can capture discriminative patterns at the cluster level, enabling it to maintain ranking quality even when individual user signals are sparse.

\begin{table}[htpb]
    \caption{Performance Comparison in Cold-Start Scenarios on KuaiRand Dataset.}
    \label{tbl_coldstart_performance}
    \centering
    \resizebox{1.0\linewidth}{!}{
        \begin{tabular}{l|ccc|ccc}
            \hline
            \textbf{Objective} & \multicolumn{3}{c|}{\textbf{KuaiRand (Cold)}} & \multicolumn{3}{c}{\textbf{KuaiRand (Warm)}} \\
            & LogLoss & AUC & GAUC & LogLoss & AUC & GAUC \\
            \hline
            LogLoss & 0.6189 & 0.7298 & 0.6718 & 0.5683 & 0.7454 & 0.6945 \\
            LogLoss + PairwiseLogistic & 0.6171 & 0.7305 & 0.6735 & 0.5670 & 0.7461 & 0.6955 \\
            LogLoss + SoftmaxCE & 0.6175 & 0.7302 & 0.6730 & 0.5674 & 0.7458 & 0.6952 \\
            LogLoss + ListCE & 0.6137 & 0.7308 & 0.6732 & 0.5662 & 0.7469 & 0.6962 \\
            JRC & 0.6145 & 0.7308 & 0.6738 & 0.5664 & 0.7475 & 0.6968 \\
            \hline
            GroupCE (proposed) & \textbf{0.6115} & \textbf{0.7320} & \textbf{0.6786} & \textbf{0.5636} & \textbf{0.7489} & \textbf{0.6986} \\
            \hline
        \end{tabular}
    }
\end{table}

We stratify users based on the number of impressions in the KuaiRand training set into cold $(\leq20)$ and warm $(20-50)$ groups, and assess model performance on the corresponding users in the test set. As shown in \autoref{tbl_coldstart_performance}, our proposed GroupCE framework consistently outperforms baselines across both user segments, with the most notable GAUC gains observed in the cold-start group. These results indicate that hierarchical clustering introduces effective user-level priors, enabling better ranking performance even with limited user history, a promising direction for addressing cold-start challenges in recommendation systems.

\subsection{Model Study (Q3)}

To gain deeper insights into the proposed model, we conduct experiments on the KuaiRand dataset and evaluate performance under different configurations. This section examines: (1) the impact of ablating individual loss components, and (2) the effect of varying codebook size and quantization depth.

\subsubsection{Ablation Study: Loss Component Analysis}
To understand the contribution of each component in our proposed framework, we conduct an ablation study by incrementally removing key loss terms from the overall training objective:

\begin{itemize}[leftmargin=10pt]
\item \textbf{w/o Hierarchical Group-Wise Ranking Losses}: This variant removes the hierarchical group-wise ranking loss $\mathcal{L}_{\text{hierarchical}}$, training the model only with the calibration losses from the original and quantized embeddings. This isolates the impact of structured negative sampling and ranking losses.
\item \textbf{w/o Quantized Auxiliary Loss}: This variant removes the auxiliary calibration loss $\mathcal{L}_{\text{logloss}}(\hat{y}^{q}, y)$, leaving only the primary calibration loss $\hat{y}$ and the hierarchical ranking loss. This examines the role of quantization-aware supervision in regularizing the user network.
\end{itemize}

\begin{table}[htpb]
    \caption{Ablation Study on Different Losses.}
    \label{tbl_ablation_loss}
    \centering
    \resizebox{0.9\linewidth}{!}{
        \begin{tabular}{l|ccc}
            \hline
            \textbf{Model Variant} & LogLoss & AUC & GAUC \\
            \hline
            w/o Hierarchical Loss & 0.5726 & 0.7518 & 0.6915 \\
            w/o Quantized Auxiliary Loss & 0.5697 & 0.7541 & 0.6939 \\
            \hline
            Full Model (GroupCE) & \textbf{0.5681} & \textbf{0.7556} & \textbf{0.6953} \\
            \hline
        \end{tabular}
    }
\end{table}

As illustrated in \autoref{tbl_ablation_loss}, the performance drop from removing either loss term highlights their complementary roles. Excluding the hierarchical ranking loss removes structured ranking optimization from hard negatives, weakening the model’s ability to learn fine-grained item ordering, especially within users, as reflected by GAUC metric. Dropping the quantized auxiliary loss removes a key regularization signal that aligns user embeddings with their discrete code representations, reducing generalization and calibration quality. Together, these components enable both effective ranking and robust user representation.

\subsubsection{Impact of Codebook Size and Quantization Depth}
To evaluate the sensitivity of our framework to quantization configuration, we vary the codebook size ($K$) and the number of quantization levels ($L$) which control the granularity and capacity of hierarchical user codes and directly influence group-wise sampling structure.

\begin{figure}[htpb]
    \centering
    \includegraphics[width=0.40\textwidth, height=0.35\textwidth]{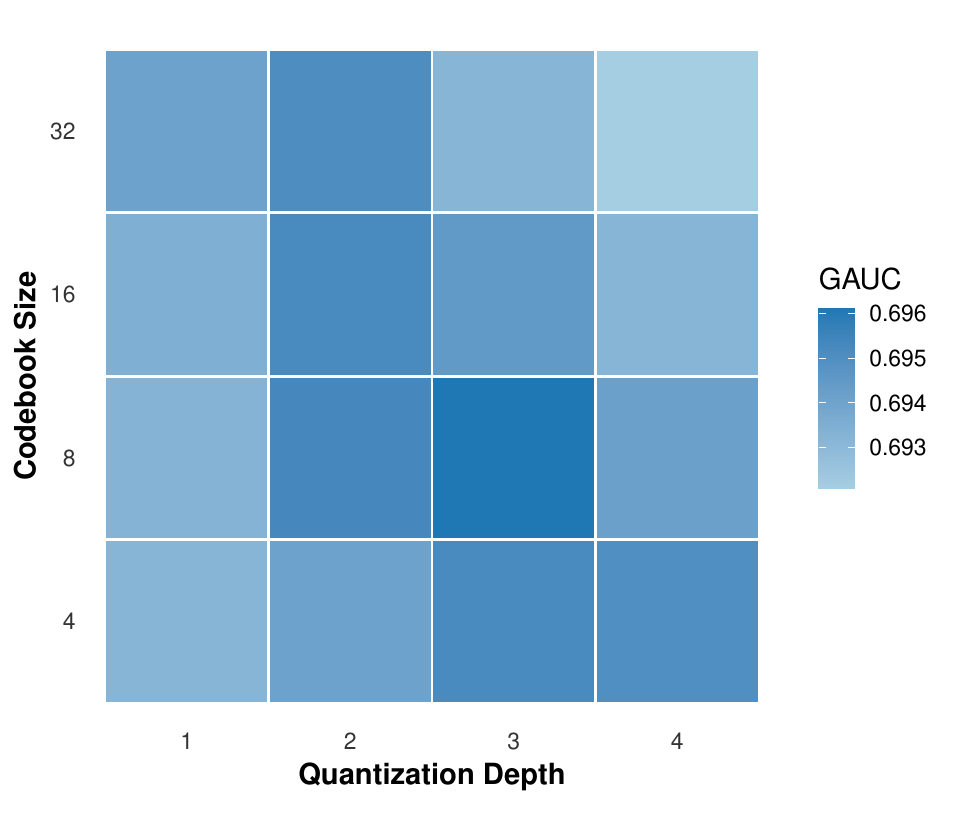}
    \caption{GAUC performance heatmap on KuaiRand dataset across different codebook sizes and quantization depths.}
    \label{fig_codebook_heatmap}
    \medskip
    \small
\end{figure}

As shown in Figure~\ref{fig_codebook_heatmap}, both codebook size and quantization depth influence performance. We observe that increasing depth consistently improves GAUC due to richer hierarchical structure and finer-grained user grouping, but further deeper quantization yields diminishing returns due to over-fragmentation and sparse supervision signals in excessively narrow groups. Similarly, moderate codebook sizes outperform extremes: small codebooks lack expressive power to form meaningful clusters while large codebooks create overly sparse groupings that reduce the effectiveness of listwise ranking. These results suggest careful hyperparameter selection is needed to balance granularity and group density for effective ranking optimization.

\section{Conclusion}
We propose a Hierarchical Group-wise Ranking Framework that enhances CTR/CVR model performance by leveraging residual vector quantization to construct hierarchical user clusters for structured, cluster-aware ranking optimization. Our approach organizes user-item pairs into meaningful semantic clusters, enabling efficient hard negative sampling while avoiding real-time context collection and expensive cross-user retrieval computations. We introduce a hierarchical listwise ranking loss to model item ordering across varying levels of user similarity and complement it with calibration objectives applied to both original and quantized embeddings. The proposed framework provides a scalable and generalizable solution for industrial recommendation systems and opens new directions for ranking optimization via quantization-based hierarchical clustering, enabling efficient learning-to-rank under limited user feedback. For future research, we plan to extend our approach to pre-ranking models that optimize ranking objectives across different sample spaces through more effective negative sampling and group-wise ranking framework.

\bibliographystyle{ACM-Reference-Format}
\bibliography{grouprank.bib}

\end{document}